\documentclass[aps,prb,reprint,superscriptaddress]{revtex4-1}

\usepackage{graphics,graphicx,amssymb,amsmath}

\begin{document}


\title{Magnetic transitions in the topological magnon insulator Cu(1,3-bdc)}

\author{R. Chisnell}
\email[]{robin.chisnell@nist.gov}
\affiliation{Department of Physics, Massachusetts Institute of Technology, Cambridge, MA 02139}
\affiliation{NIST Center for Neutron Research, Gaithersburg, MD 20899}

\author{J. S. Helton}
\altaffiliation{Current Address: Department of Physics, United States Naval Academy, Annapolis, MD 21402}
\affiliation{NIST Center for Neutron Research, Gaithersburg, MD 20899}

\author{D. E. Freedman}
\affiliation{Department of Chemistry, Massachusetts Institute of Technology, Cambridge, MA 02139}
\affiliation{Department of Chemistry, Northwestern University, Evanston, IL 60208}

\author{D. K. Singh}
\affiliation{Department of Physics and Astronomy, University of Missouri, Columbia, MO 65211}

\author{F. Demmel}
\affiliation{ISIS Facility, Rutherford Appleton Laboratory, Chilton, Didcot, OX11 0QX, Oxfordshire, UK}

\author{C. Stock}
\altaffiliation{Current Address: School of Physics and Astronomy and Centre for Science at Extreme Conditions, University of Edinburgh,
Edinburgh EH9 3FD, UK}
\affiliation{ISIS Facility, Rutherford Appleton Laboratory, Chilton, Didcot, OX11 0QX, Oxfordshire, UK}

\author{D. G. Nocera}
\affiliation{Department of Chemistry, Massachusetts Institute of Technology, Cambridge, MA 02139}
\affiliation{Department of Chemistry and Chemical Biology, Harvard University, Cambridge, MA 02138}

\author{Y. S. Lee}
\email[]{youngsl@stanford.edu}
\affiliation{Department of Physics, Massachusetts Institute of Technology, Cambridge, MA 02139}
\affiliation{Department of Applied Physics, Stanford University, Stanford, California 94305, USA}
\affiliation{Stanford Institute for Materials and Energy Sciences, SLAC National Accelerator Laboratory, Menlo Park, California 94025, USA}

\date{\today}

\begin{abstract}
Topological magnon insulators are a new class of magnetic materials that possess topologically nontrivial magnon bands.  As a result, magnons in these materials display properties analogous to those of electrons in topological insulators.  Here we present magnetization, specific heat, and neutron scattering measurements of the ferromagnetic kagome magnet Cu(1,3-bdc).  Our measurements provide a detailed description of the magnetic structure and interactions in this material and confirm that it is an ideal prototype for topological magnon physics in a system with a simple spin Hamiltonian.
\end{abstract}


\maketitle

\section{Introduction}

In systems where quantum particles are confined to move in reduced dimensions, lattice geometry and particle interactions can drive the emergence of a rich variety of novel behaviors.  A canonical example is the quantum Hall effect, which is produced by applying a large magnetic field to a two-dimensional (2D) gas of electrons or charged quasiparticles \cite{Klitzing}.  Some systems exhibit quantum Hall physics without applied magnetic fields due to their inherently topological band structures, as demonstrated by Haldane \cite{Haldane}.  Many materials host topological band structures as a result of strong spin-orbit coupling, including the well known topological insulators.  Discoveries of these materials have driven much condensed matter physics research \cite{HasanReview,QiReview}.  Recent theoretical work has been most interested in 2D systems whose band structures (1) include at least one band that is dispersionless in energy(flat) and (2) are topologically nontrivial.  These studies hope to achieve the fractional quantum Hall effect without an externally applied magnetic field \cite{RoyReview}.  Flat bands are of particular interest because the large number of states with degenerate kinetic energy allows the particle interactions to dominate the behavior, which can result in the emergence of novel strongly correlated phenomena.  A number of recent theoretical works have proposed models that use flat topological bands to produce the fractional quantum Hall effect \cite{Tang,Sun,Neupert}; however, these models require tuning of multiple parameters, which is not always possible in real materials. 

Topological band structures can be found in a variety of different systems, not only those with electronlike quasiparticles.  In fact, many systems with bosonic quasiparticles display topological band structures.  For example, topological photon modes can be realized in photonic crystals \cite{Raghu,Wang,Lu}.  Theoretical proposals for realizations of bosonic systems with bands that are both topologically nontrivial and dispersionless include dipolar molecules trapped in an optical lattice \cite{YaoLukin} and photonic lattices \cite{Petrescu} based on the interaction between photons and arrays of superconducting circuits \cite{Koch}, although experimental confirmation has yet to be demonstrated.  Additionally, topological magnon band structures can be found in insulating ferromagnets called topological magnon insulators (TMIs) \cite{Zhang}.  In these materials, the magnon band structure includes gapless, nondissipative edge modes within a bulk band gap, analogous to the electronic band structure of electronic topological insulators.

We have recently shown the existence of topological magnon bands in the kagome lattice compound Cu[1,3-benzenedicarboxylate(bdc)] \cite{bdcInelastic}, the first realization of a 2D TMI.  Cu(1,3-bdc) is a metal-organic hybrid material featuring $S$ = 1/2 Cu$^{2+}$ ions arranged on a geometrically perfect kagome lattice \cite{Nytko}.  Adjacent kagome planes are well separated by large organic (1,3-bdc) molecules, as shown in Fig. 1(a).  The long interplane coupling pathway suggests that the magnetic behavior should be quasi-two-dimensional and the absence of other metal ions leads to less chance of disorder in the kagome plane.  These three properties -- the structurally perfect kagome lattice, the absence of other species of metal ion, and the weak interlayer coupling -- make Cu(1,3-bdc) an ideal model material for examining fundamental physics with a simple spin Hamiltonian. 

Magnons in Cu(1,3-bdc) display a number of novel properties.  For example, due to the kagome geometry \cite{Bergman}, one of the topologically nontrivial magnon bands is also nearly dispersionless.  Additionally, magnons in this material display a magnon Hall effect \cite{ThermalHallTheory,ThermalHallTheory2}, as confirmed by thermal Hall measurements \cite{Hirschberger}.  The magnon Hall effect may find applications in the field of spintronics \cite{Spintronics}, as has been proposed for a number of pyrochlore ferromanets that also display a magnon Hall effect\cite{Onose,Ideue}.  Finally, as a realization of the TMI state, magnons in Cu(1,3-bdc) are expected to display topologically protected chiral edge modes\cite{Zhang}, making Cu(1,3-bdc) an experimental system that could be used to test ideas regarding the use of edge magnons to manipulate skyrmions \cite{Pereiro}, as well as theoretical predictions of hybridization of edge modes \cite{Mook}.

For the many possible applications of Cu(1,3-bdc), a thorough understanding of the magnetically ordered state and of the ordering transition is essential.  Our previous neutron scattering measurements demonstrated the existence of a long-range magnetic ordering transition where spins within each kagome layer are ordered ferromagnetically, while spins in neighboring kagome planes are oriented antiferromagnetically \cite{bdcInelastic}.  Previous magnetic and specific heat measurements are consistent with a ferromagnetic ordering transition near 1.8 K, despite a negative Curie-Weiss temperature that suggests antiferromagnetic nearest-neighbor interactions \cite{Nytko}.  In contrast, muon spin resonance ($\mu$SR) measurements suggested that below the transition temperature the fluctuation rate of the spins was slowed, but that there was no long-range ordering of the moments \cite{bdcMuon}.  In this report, we present magnetization, specific heat, and neutron scattering measurements of Cu(1,3-bdc) and examine the nature of the magnetically ordered state and the magnetic phase transition.

\begin{figure}
\includegraphics[width=8.5cm]{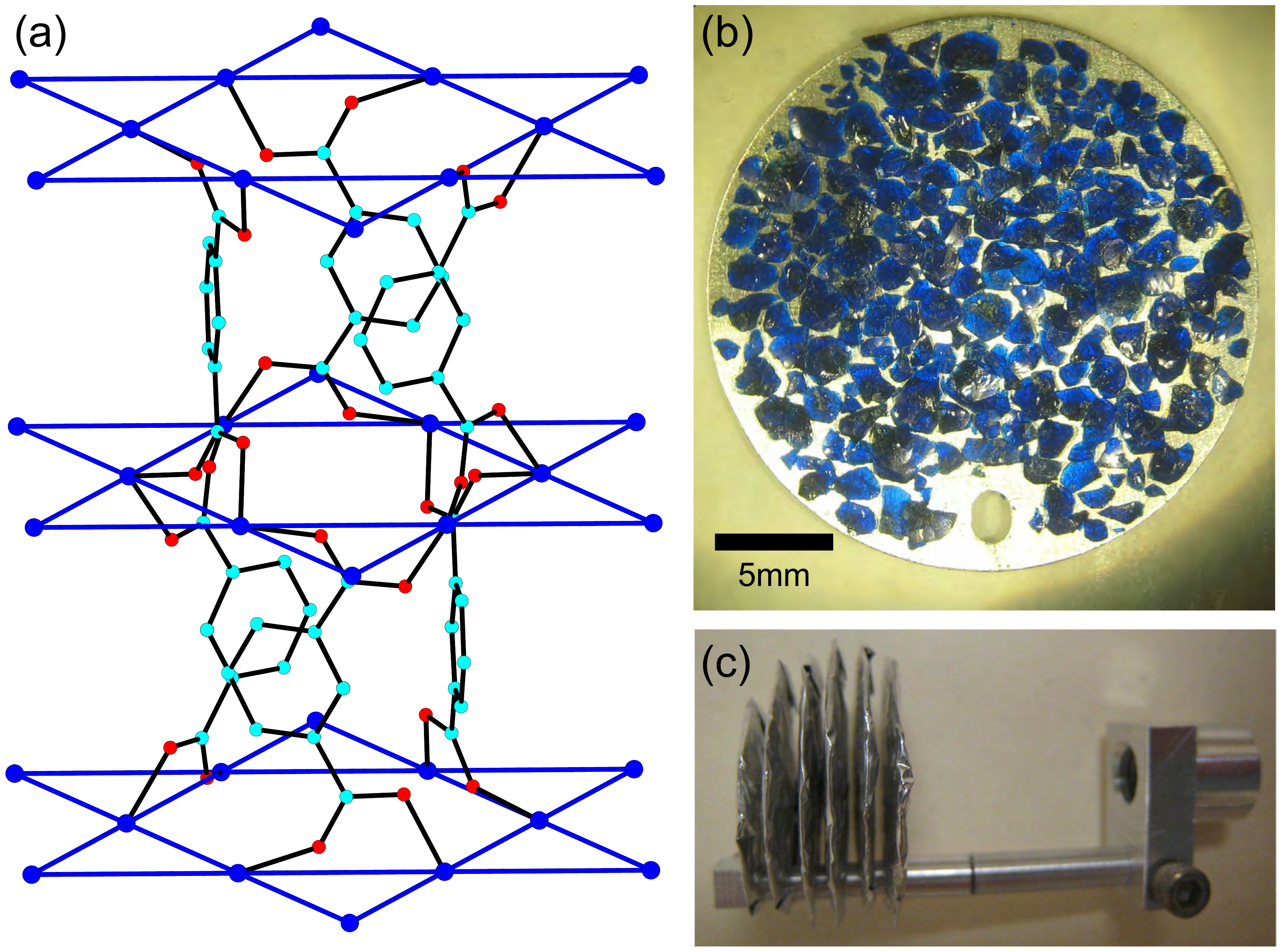} \vspace{0mm}
\caption{(a) Crystal structure of Cu(1,3-bdc).  Copper ions (blue) form kagome lattice layers separated by benzene dicarboxylate(bdc) molecules containing oxygen (red), carbon (cyan), and hydrogen (not shown).  (1,3-bdc) molecules not connected to the center hexagons and all hydrogen atoms have been removed for clarity.  (b) One plate of the $\vec{c}$-axis-aligned sample used for neutron scattering measurements. (c) Assembled neutron scattering sample with six parallel plates.}
\end{figure}

\section{Experimental Details}

Protonated and deuterated crystals of Cu(1,3-bdc) were grown under hydrothermal conditions using a similar procedure to that reported by Nytko $\emph{et al.}$ \cite{Nytko}.  To increase crystal size, the reaction rate was slowed by using dilute nitric acid in place of water.  Cu(1,3-bdc) is produced by reaction of Cu(OH)$_2$ with isophthalic acid (1,3-bdcH$_2$).  Since deprotonation of the isophthalic acid is likely the rate-determining step, the reaction rate can be slowed by lowering the pH and therefore reducing the concentration of deprotonated 1,3-bdc$^{2-}$ in solution.

Reactions were carried out in 23 and 125 mL PTFE-lined pressure vessels.  To synthesize deuterated crystals, 23-mL (125-mL) liners were charged with 145 mg (435 mg) Cu(OH)$_{2}$, 250 mg (750 mg) deuterated isophthalic-d4 acid (1,3-bdcH$_2$), and 8.5 g (25.5 g) 2 $\%$ nitric acid in water, and placed into steel hydrothermal bombs.  Bombs were heated to 150 $^{\circ}$C and maintained at this temperature for 21 days, removed from the furnace at temperature and cooled in air, resulting in clusters of crystals of Cu(1,3-bdc) forming at the bottoms of the liners.  Crystal clusters were washed in deionized water and dried in air.  Single crystal pieces were manually separated from clusters under a microscope, resulting in crystals with a mass typically 0.1 to 1 mg but up to 3 mg.  Protonated crystals were synthesized using a similar procedure except using nondeuterated isophthalic acid.  For protonated crystals, a lower growth temperature of 130 $^\circ$C was used in order to further increase crystal size.

In order to assemble a sample with enough total mass for neutron scattering measurements, we partially coaligned $\approx$2000 individual deuterated crystals.  Single crystal pieces of Cu(1,3-bdc) form as flat flakes with the $\vec{c}$ axis perpendicular to the plane of the flat face.  Therefore, the $\vec{c}$ axes were aligned by arranging crystals on fllat aluminum plates.  The orientation of the kagome plane of each crystal was not aligned and is assumed to be random.   Figure 1(b) shows one such plate.  Crystals were attached to both sides of the plates using Fomblin Y oil and secured with aluminum foil. Six plates were held parallel to each other to create a $\vec{c}$-axis-aligned sample with total mass 1 g, shown in Fig. 1(c).

Magnetization measurements were performed on both protonated and deuterated single crystal samples of Cu(1,3-bdc) using a Quantum Design Magnetic Property Measurement System (MPMS).  Magnetization was measured as a function of field for fields up to 7 T at temperatures $T$ = 1.8 K, 5 K, and 30 K, and as a function of temperature over the range 1.8 K to 350 K at a number of applied fields ranging from 2 mT to 5 T.  Measurements were performed with the field applied parallel to and perpendicular to the kagome plane.  Low-temperature ($T <$ 10 K) measurements were performed under both field-cooled (FC) and zero-field-cooled (ZFC) conditions.  The specific heat of a protonated single crystal sample of Cu(1,3-bdc) was measured using a Quantum Design Physical Property Measurement System (PPMS).  Measurements were performed in applied fields up to 14 T with the field applied parallel to and perpendicular to the kagome plane.

Neutron diffraction and elastic scattering measurements were performed on the deuterated $\vec{c}$-axis-aligned sample described above using the triple-axis spectrometer SPINS at the NIST Center for Neutron Research.  Elastic measurements were done at zero magnetic field with the sample in a He-4 cryostat using neutrons of energy $E = 3$ meV with the configuration guide-$80'$-$80'$-open.  Be and BeO filters were placed before and after the sample, respectively.  Diffraction measurements were done with the sample in a dilution insert in a 10-T magnet using neutrons of initial energy $E_i = 5$ meV.  Measurements were perfomed in two-axis mode with the configuration guide-$80'$-$80'$, with Be filters placed before and after the sample.  The sample was oriented so that the aligned $\vec{c}$ axis was in the scattering plane and the magnetic field was applied parallel to the kagome plane.  Unless otherwise noted, reported diffraction data were measured after application of a magnetic field at low temperatures in order to suppress the superconductivity of the aluminum sample holder and ensure thermal equilibrium of the sample.

Inelastic neutron scattering measurements were performed on a deuterated powder sample of Cu(1,3-bdc) on the Iris time-of-flight spectrometer at the ISIS facility at Rutherford Appleton Laboratory.  An aluminum can was filled with 3.9 g of powder and helium exchange gas and placed in a dilution refrigerator.  A final neutron energy of 7.38 meV was selected, giving an energy resolution of $\approx$70 $\mu eV$ FWHM.

\section{Results and Discussion}

\subsection{Crystal Characterization}

\begin{figure}
\includegraphics[width=8.6cm]{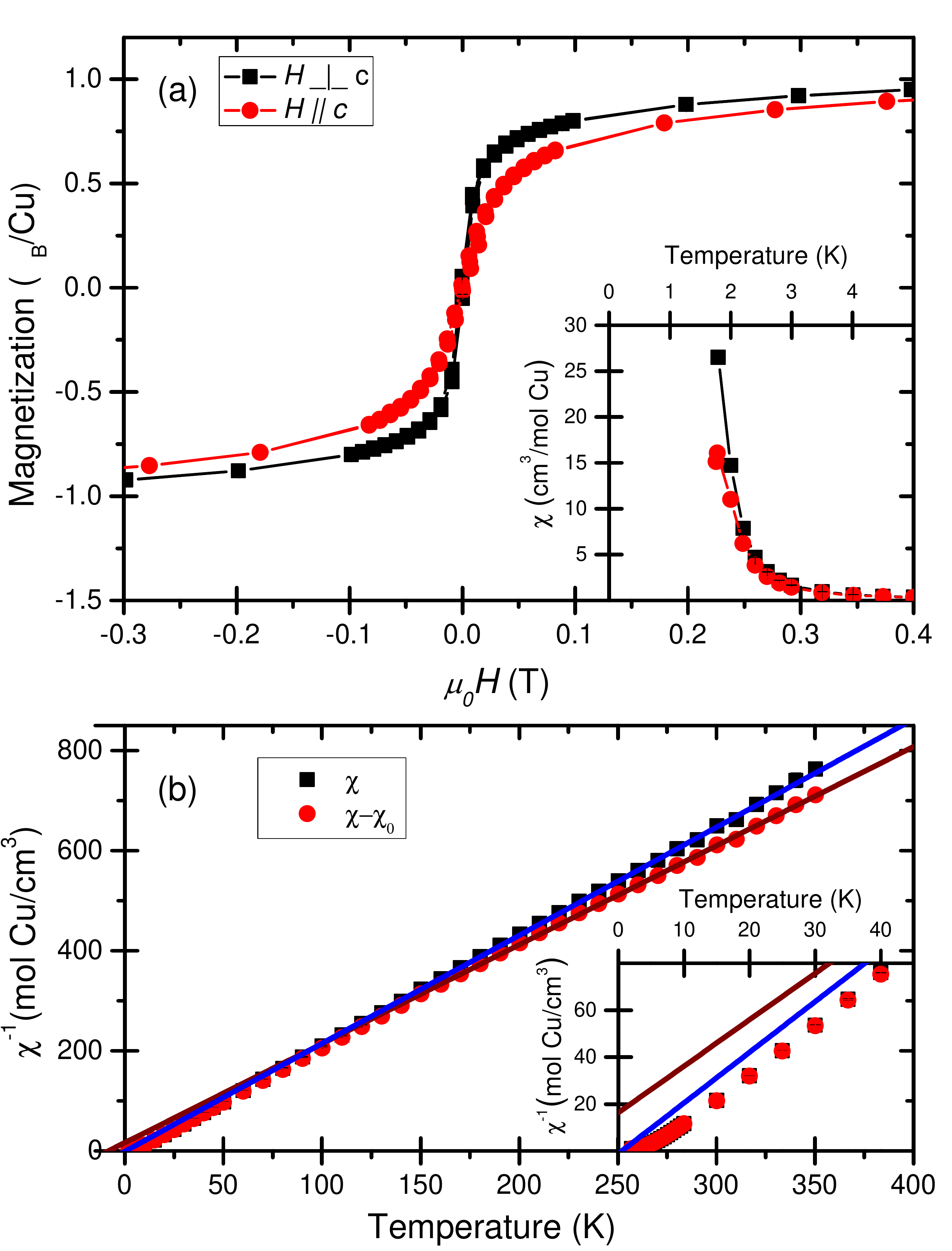}
\caption{(a) Magnetization as a function of applied magnetic field at 1.8 K.  (Inset) Susceptibility as a function of temperature at an applied field of $\mu_0H=0.01$ T.  The field was applied both parallel ($H\bot c$) and perpendicular ($H||c$) to the kagome plane.  Lines are guides for the eye.  (b) Inverse susceptibility measured with a $\mu_0H$ = 0.5 T field applied parallel to the kagom\'e plane.  Lines are Curie-Weiss fits over the range 150 K to 350 K to the measured susceptibility (blue) and the susceptibility corrected for the molecular diamagnetic contribution (red).  The inset shows the low-temperature region.  Demagnetization corrections have been applied as described in the text.}
\end{figure} 

\begin{figure}
\includegraphics[width=8.6cm]{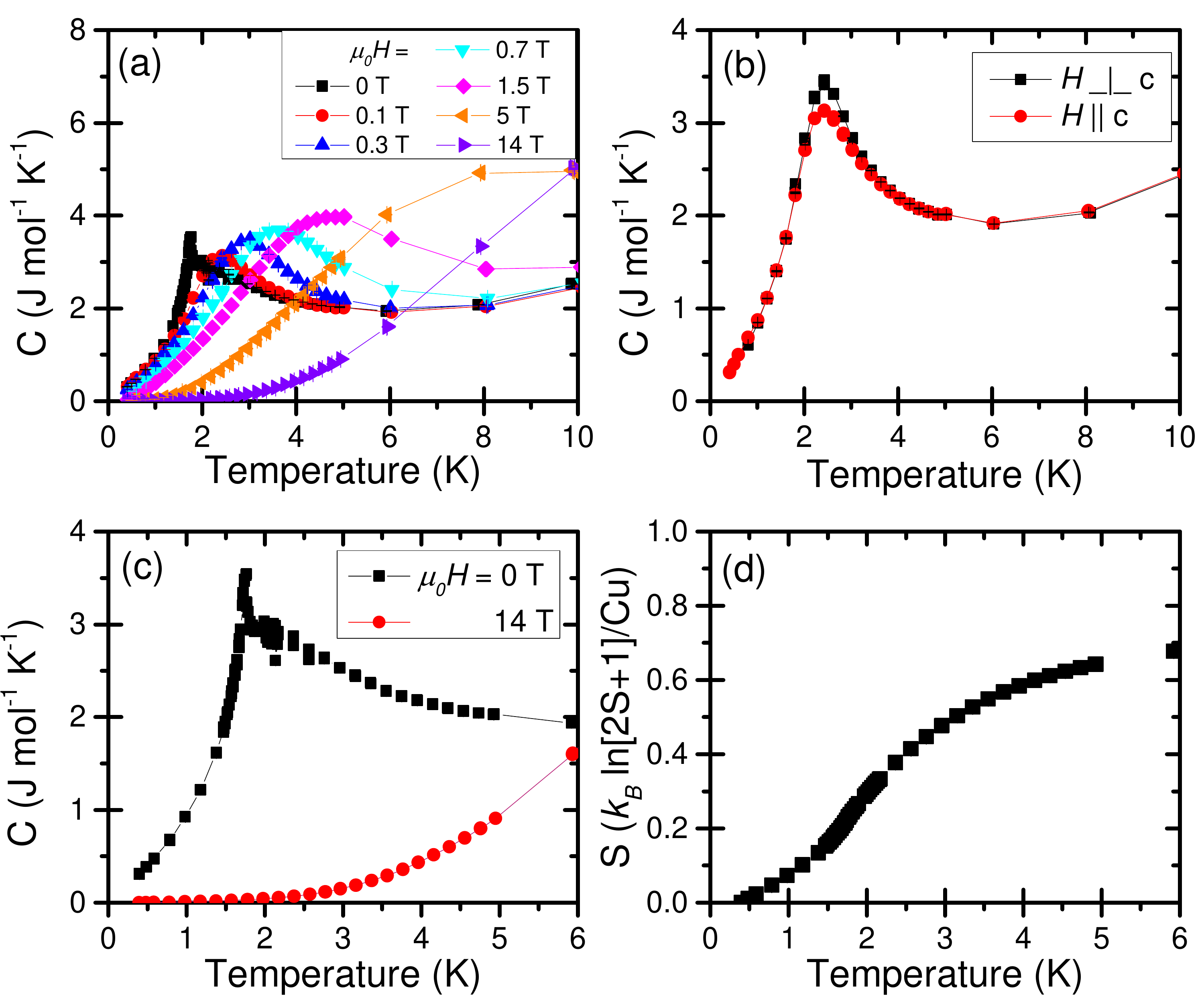}
\caption{Temperature dependence of specific heat of Cu(1,3-bdc).  (a) Specific heat under applied fields up to 14 T with the field perpendicular to the kagome plane.  (b) Comparison of specific heat with the field applied parallel to the kagome plane ($H\bot c$) and perpendicular to the kagome plane ($H||c$) with $\mu_0H=0.1$ T.  (c) Specific heat measured at zero field and a 14-T field applied perpendicular to the kagome plane.  The 14-T data are assumed to be dominated by the phonon contribution with negligible magnetic contribution in the temperature range $T\leq5$ K.  (d) Magnetic entropy starting at $T$ = 0.4 K calculated from the data in (c).}
\end{figure}

Figure 2(a) shows the magnetization of Cu(1,3-bdc) as a function of applied field at $T=1.8$ K and the susceptibility $\chi$ approximated as $M/H$ measured at a field of $\mu_0H$ = 0.01 T, for fields applied both parallel ($H\bot c$) and perpendicular ($H||c$) to the kagome plane.  The magnetization is easily saturated by fields of $\approx 2$ T at $T=1.8$ K for both field directions.  The susceptibility is enhanced, and saturation is reached at a lower field when the field is applied parallel to the kagome plane.  The enhancement in susceptibility persists above the transition to temperatures as high as $T\approx 4$ K.  Very little hysteresis is observed, with a coercive field less than 2 mT.  No difference is observed between ZFC and FC measurements.  These results are consistent with previous measurements performed on powder \cite{Nytko} and single crystal \cite{bdcMag} samples.

An anisotropic magnetization signal is expected because the flat plate shape of the measured crystal results in an anisotropic demagnetizing field.  Data in Fig. 2 have been corrected for the demagnetizing field follwing $H = H_0 - 4\pi N M$, where $H_0$ and $H$ are the applied field and total field at the sample, respectively, in Oe, $N$ is the demagnitization factor, and $M$ is the magnetization in emu/cm$^3$.  The sample shape is approximated as a short cylinder\cite{demag} with 3 mm diameter and 0.2 mm height, giving $N(H\bot c) = 0.065$ and $N(H||c) = 0.87$.  The difference observed between the two measurements is reduced, but not eliminated by this correction.   

At low applied fields, no difference in magnetic behavior is observed between protonated and deuterated samples.  At high fields, slight differences were observed in the saturation value of the magnetization, which was 1.166(9) $\mu_B$/Cu with $H\bot c$ and 1.056(5) $\mu_B$/Cu with $H||c$ for the protonated sample and 1.109(3) $\mu_B$/Cu with $H\bot c$ and 1.075(8) $\mu_B$/Cu with $H||c$ for the deuterated sample.  This confirms that deuterium substitution has little effect on the magnetic behavior of Cu(1,3-bdc).

Figure 2(b) shows the inverse susceptibility of a deuterated crystal measured at $\mu_0H$ = 0.5 T with the field applied parallel to the kagome plane.  This plot shows both the raw data ($\chi$) and the data corrected for the molecular diamagnetism of the sample ($\chi$-$\chi_0$) by use of Pascal's constants\cite{PascalsConstants}.  This correction is done to isolate the susceptibility due to the copper spins.  Both sets of data were fit to a Curie-Weiss function over the temperature range 150 K $\leq T \leq$ 350 K.  Surprisingly, the raw inverse susceptibility data are better fit by a linear model than the data corrected for the sample's diamagnetism.  The data also deviate less from the fit line at lower temperatures.  The fit to the raw data gives a Curie-Weiss temperature of $\Theta_{CW}$ = 0.5 $\pm$ 0.3 K, while the fit to the corrected data gives $\Theta_{CW}$ = $- 8.3$ $\pm$ 0.3 K.  A positive Curie-Weiss temperature suggests a ferromagnetic nearest-neighbor exchange coupling, while a negative temperature suggests antiferromagnetic coupling.  This sign discrepancy makes it difficult to determine the coupling from the inverse susceptibility data.  We therefore abandon Curie-Weiss fits as an accurate characterization of the Cu(1,3-bdc) spin Hamiltonian.  Our previously reported inelastic neutron scattering measurements \cite{bdcInelastic} demonstrated that nearest-neighbor coupling is ferromagnetic with $J=0.6$ meV ($J/k_B=7$ K).

The temperature dependence of the specific heat of Cu(1,3-bdc) is shown in Fig. 3.  Figure 3(a) displays the data measured at a number of fields with the field applied perpendicular to the kagome plane.  In zero applied field, a peak is observed at $T$ = 1.77 K, which is similar to that observed in previous measurements \cite{Nytko} and consistent with the onset of magnetic order at that temperature.  With increasing applied field, the peak broadens and shifts to higher temperatures.  Figure 3(b) compares measurements with a field of $\mu_0H=0.1$ T applied in the two different directions.  No difference is observed at temperatures far away from the peak in the specific heat.  At temperatures near the peak, the specific heat is enhanced when the field is applied parallel to the kagome plane.  This effect is suppressed at high fields and is no longer detectable for fields greater than 5 T.  A 14-T field shifts magnetic scattering to higher temperatures, and reveals that the specific heat is dominated by the magnetic contribution below $\approx 5$ K, as shown in Fig 3(c).

Figure 3(c) shows that there is a significant magnetic contribution to the specific heat at temperatures well above the transition temperature.  We isolate the magnetic contribution to the zero-field specific heat by subtracting the specific heat measured at $\mu_0H=14$ T, and use this to calculate the magnetic entropy, which is shown in Fig. 3(d).  The magnetic entropy released in zero field below the transition temperautre $T$ = 1.77 K down to the lowest measured temperature of $T$ = 0.4 K was found to be only 24 \% of $k_B$ln(2) per spin, while about 40 \% is released between 5 K and the transition temperature.  At higher temperatures, the approximation of the 14 T data as nonmagnetic becomes invalid, but the zero-field magnetic contribution to the specific heat is nonzero at least up to $T$ = 6.5 K, where the zero-field and 14 T data intersect.  Some entropy may also be lost at temperatures lower than could be measured, but it is clear that a large fraction of the magnetic entropy is released at temperatures well above the transition temperature.

\subsection{Magnetic Structure}

Longitudinal scans and $\theta$ scans were performed through the (0 0 $L$) Bragg peaks for $L = 1$ to 6 at the base temperature of 70 mK.  The (0 0 $L$) structural peaks are forbidden for odd values of $L$ because there are two identical kagome layers per unit cell.  Bragg peaks were observed at all six measured positions.  Peaks at (0 0 $L$) for odd values of $L$ were quickly suppressed by the application of a magnetic field.  These peaks also disappeared above 1.8 K.  Figures 4(a) and 4(b) show representative scans through (0 0 1).  The application of a magnetic field also enhanced the intensity of the peak at (0 0 4), as shown in Fig. 4(c) and 4(d).  

In contrast with the $\mu$SR result \cite{bdcMuon}, our observation of the emergence of magnetic Bragg peaks below 1.77 K is a clear sign of a transition to a state with long-range magnetic ordering.  The magnetic peaks at (0 0 $L$) for odd values of $L$, where structural peaks are forbidden, demonstrate the existence of antiferromagnetic ordering between neighboring kagome planes.  The suppression of these peaks with magnetic field along with the growth of the (0 0 4) peak shows that the material is easily pushed into a fully spin polarized state, which is consistent with the easily saturated magnetization (Fig. 2).  This easy saturization, combined with our observation of magnons consistent with ferromagnetic coupling \cite{bdcInelastic}, demonstrates that spins within each kagome plane are ordered ferromagnetically. 

\begin{figure}
\includegraphics[width=8.5cm]{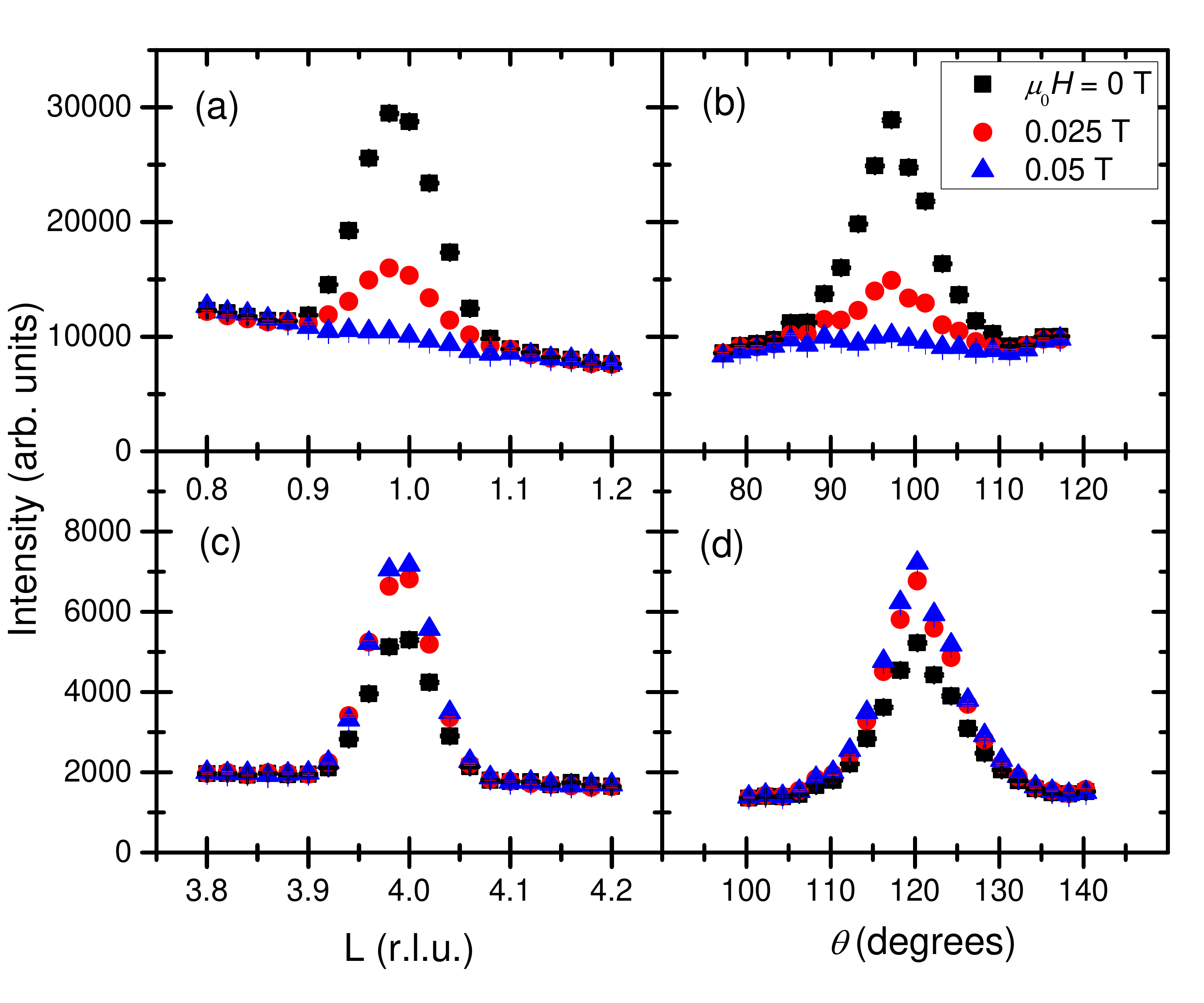} \vspace{0mm}
\caption{\label{Figure1} Neutron diffraction scans through the (a),(b) (0 0 1) and (c),(d) (0 0 4) Bragg peak positions at $T=70$ mK.  The application of a magnetic field quickly suppresses the Bragg peak at (0 0 1) while increasing intensity at the (0 0 4) peak position.  }
\end{figure}

To investigate the zero-field ordered magnetic state, we examined the integrated intensities of the measured Bragg peaks.  The scattered intensity from a magnetically ordered crystal is proportional to the component of the spin that is perpendicular to the momentum $\vec{Q}$ \cite{Shirane}.  If the spins point parallel to the $\vec{c}$ axis, there will be no magnetic scattering at (0 0 $L$) positions.  Therefore, the ground-state spins must have some component parallel to the kagome plane for the (0 0 $L$) magnetic peaks to be observed.

The integrated intensity of a $\theta$ scan through a nuclear (magnetic) Bragg reflection in a typical neutron diffraction experiment is proportional to
\begin{align}
\mathcal{I}\propto\frac{\left|\mathbf{F}(hkl)\right|^{2}}{\sin2\theta}
\end{align}
where $\mathbf{F}$ is the static nuclear (magnetic) structure factor and 2$\theta$ is the angle between the incident and diffracted beam \cite{Shirane}.

A $\theta$ scan rotates the sample through the Bragg reflection.  Integrating over this scan accounts for the horizontal divergence of the beam and for the mosaic width of the sample in the scattering plane.  The SPINS spectrometer also includes a vertically focusing monochromator, so the incident beam includes neutrons with some component of their momentum perpendicular to the scattering plane.  This beam divergence combined with the broad sample mosaic -- due to imperfect alignment of the individual crystals -- results in a decrease in measured intensity at higher values of $\vec{\left|Q\right|}$, because of the finite detector height.

To account for the effect of the vertical beam divergence on measured Bragg intensity, we calculated the fraction of the scattered beam that would be incident on the detector as a function of $\vec{\left|Q\right|}$,
\begin{align}
\iint P_{mono}(k_{i}^{z})*P_{sample}(\Delta k^{z},\vec{\left|Q\right|})*P_{det}(k_{f}^{z})\; \mathrm{d}k_{i}^{z}\mathrm{d}k_{f}^{z}
\end{align}
where $P_{mono}$ is the distribution of neutrons with initial vertical component of momentum (k$_{i}^{z}$) leaving the monochromator, $P_{sample}$ is the probability that a scattered neutron has its vertical component of momentum changed by $\Delta k^{z}=k_{f}^{z}-k_{i}^{z}$, and $P_{det}$ selects the neutrons with the correct $k_{f}^{z}$ to be incident upon the detector.

\begin{figure}
\includegraphics[width=8.5cm]{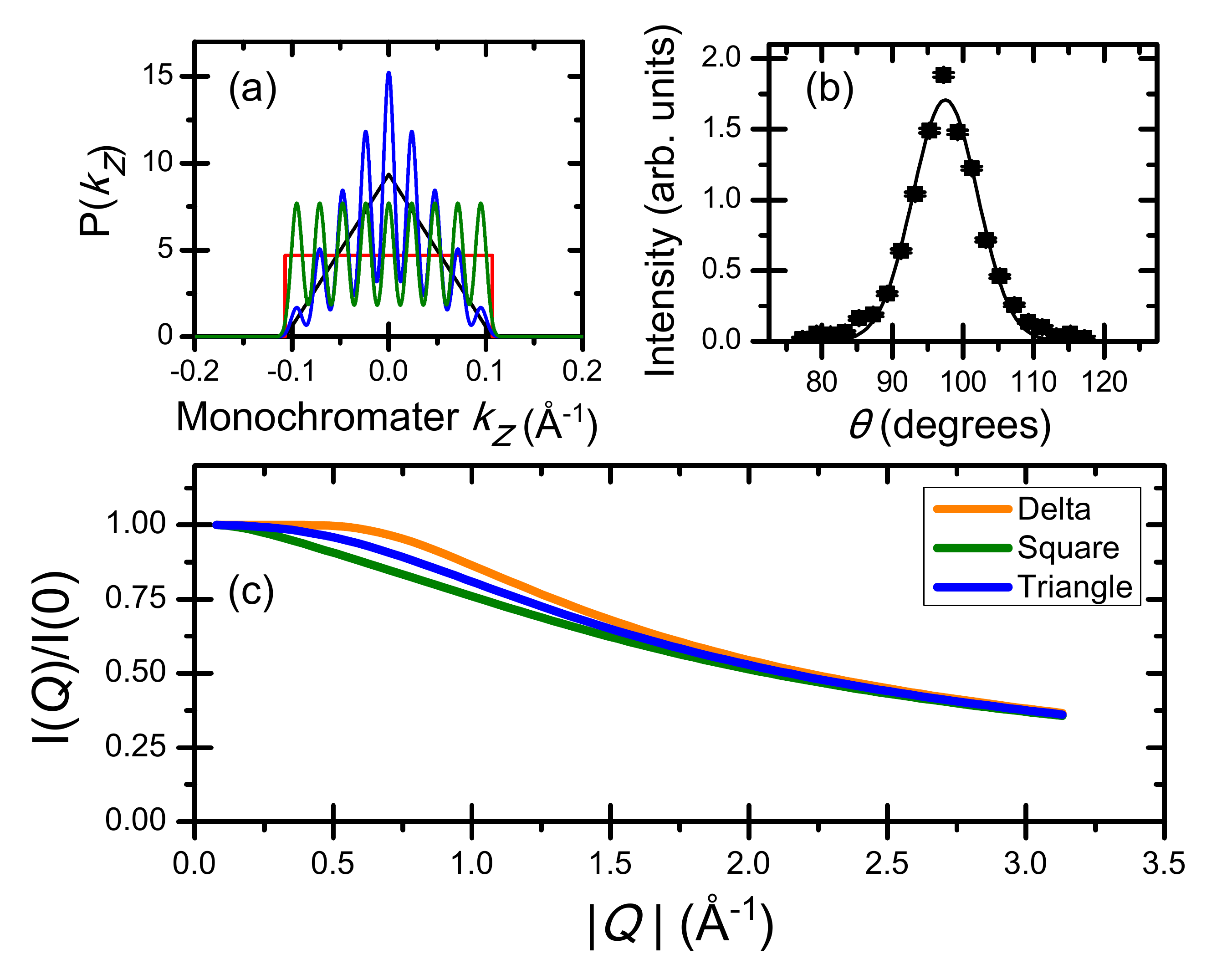} \vspace{0mm}
\caption{Decrease of measured Bragg intensity due to vertical beam divergence.  (a) Models of probability distribution of the $z$-component of the initial neutron momentum due to focusing monochromator.  (b) Background-subtracted $\theta$-scan through the (0 0 1) Bragg peak showing mosaic width of the $\vec{c}$-axis-aligned crystal sample in the scattering plane.  A $\theta$-scan taken at an applied field of $\mu_{0}H=0.5$ T was used as background.  Line is a fit to a Gaussian with width $\sigma = 4.6^{\circ}$  (c) Predicted scaling of measured intensity due to the vertical divergence as a function of $\vec{\left|Q\right|}$, calculated as described in the text.}
\end{figure}

Figure 5(a) shows different models that were assumed for $P_{mono}$.  The SPINS monochromator consists of nine pyrolytic graphite blades, each with a mosaic width of 30$'$.  We modeled the incoming vertical distribution as nine sources each with a Gaussian distribution about a different mean value of $k_{i}^{z}$.  We tried both equal weighting and a triangular weighting of these nine Gaussians.  We also tried models that ignored the details of the monochromator and just treated it as a source with finite size.  With these models we also tried both a square and a triangular distribution.  We also considered the case of perfect vertical collimation, where $P_{mono}$ is modeled as a $\delta$ function and all neutrons are assumed to have initial vertical momentum $k_{i}^{z}=0$.

Figure 5(b) shows a $\theta$ scan at the (0 0 1) Bragg position at zero applied field.  The scan taken at $\mu_{0}H=0.5$ T was used as background and subtracted from the zero-field scan.  This scan provides a measurement of the mosaic width of the sample in the scattering plane.  The mosaic is approximately Gaussian with width 4.6$^{\circ}$.  The $\vec{c}$-axis-aligned crystals that make up the measured sample have no preferred orientation in the plane perpendicular to $\vec{c}$.  Therefore, the mosaic perpendicular to the scattering plane should be similar to the mosaic in the scattering plane.  We model the vertical mosaic as a Gaussian with width 4.6$^{\circ}$.  $P_{sample}$ depends both on the mosaic width and on the momentum of the Bragg reflection.  A neutron scattering from a crystal inclined by an angle $\phi$ from the scattering plane will have its vertical component of momentum changed by $\Delta k^{z}=-Q^{z}=-\vec{\left|Q\right|}\sin(\phi)$.  Using this relation we convert the mosaic distribution to a $\vec{\left|Q\right|}$-dependent distribution of $\Delta k^{z}$.

$P_{det}$ was assumed to be a square distribution to reflect the finite size of the detector.  It depends only on the size of the detector, the sample to detector distance, and the energy of the diffracted neutrons.  Figure 5(c) shows the calculated intensity as a function of $\vec{\left|Q\right|}$ for our different models.  There was very little difference between the square distribution of monochromater blades and the overall square distribution, and very little difference between the two triangular distributions as well.  At low values of $\vec{\left|Q\right|}$, the calculated intensity is slightly dependent on the chosen model of $P_{mono}$.  At higher $\vec{\left|Q\right|}$, the calculation becomes independent of the model of $P_{mono}$, but more dependent on the Gaussian width used in the model of $P_{sample}$.  To account for these differences, we included an error bar of 10 $\%$ of the calculated value in calculations using this predicted scaling.  

\begin{figure}
\includegraphics[width=8.5cm]{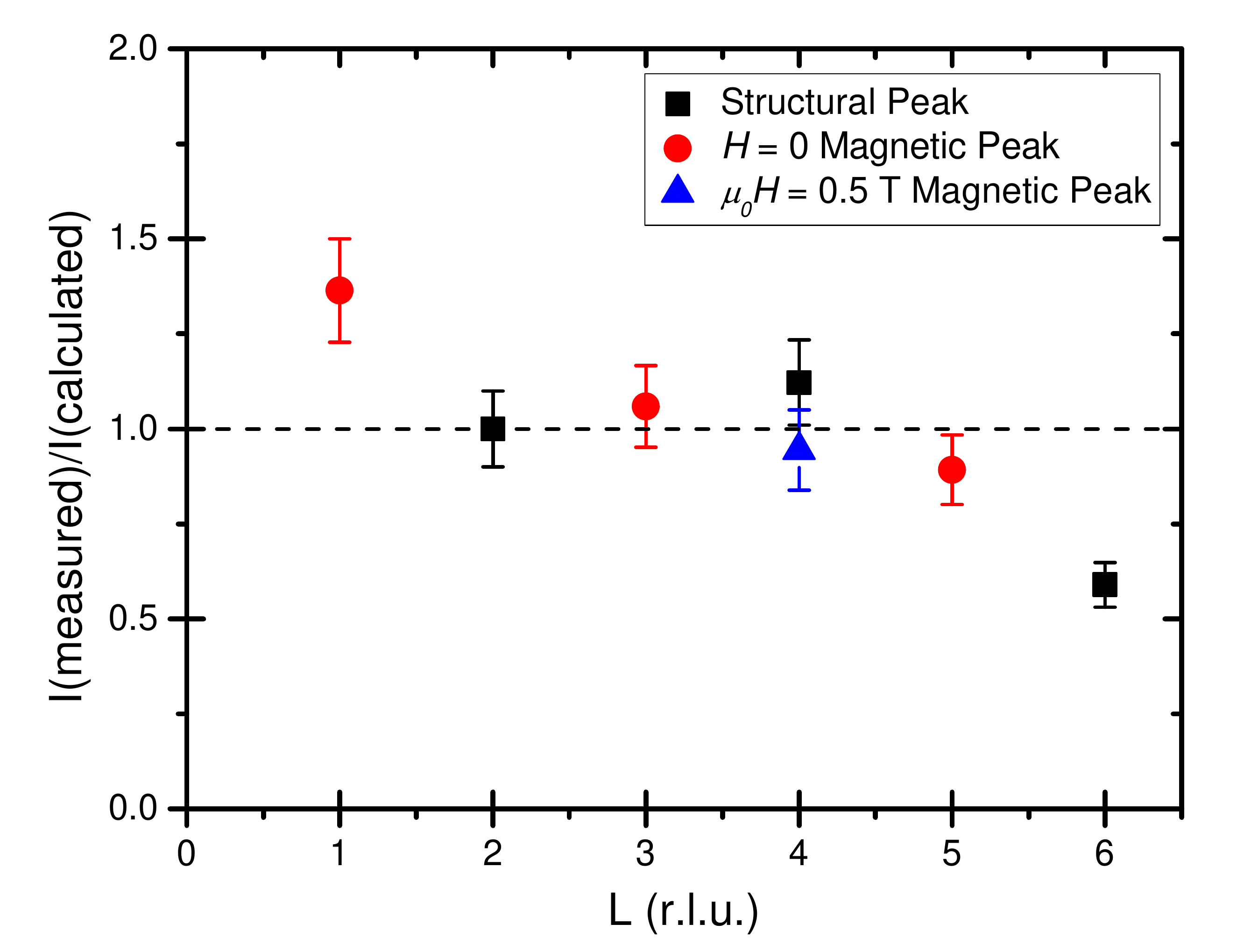} \vspace{0mm}
\caption{Integrated intensities of $\theta$ scans at (0 0 $L$) Bragg peaks as a fraction of calculated intensity, normalized to the (0 0 2) Bragg peak intensity.  The dashed line at 1.0 is a guide for the eye, and indicates the value where measured and calculated intensities are equal.}
\end{figure}

We calculated the expected intensity for each of the measured Bragg peaks,
\begin{align}
\mathcal{I}\propto\frac{\left|\mathbf{F}(00\mathrm{L})\right|^{2}}{\sin2\theta}\times \mathrm{V(00L)}
\end{align}
Where V(00$L$) is the calculated fraction of total intensity that will be measured due to the vertical divergence of the beam.  For odd values of $L$, $\mathbf{F}$(00$L$) was calculated assuming spins were ferromagnetically ordered within each kagome plane and antiferromagnetically ordered between planes, and that they were confined to the kagome plane.  We also calculated the intensity for the structural peaks and for the field-induced ferromagnetic peak at (0 0 4).  For this we assumed full ferromagnetic ordering again with the spins confined to the kagome plane.  The magnetic form factor was assumed to be the free Cu$^{2+}$ ion form factor\cite{NeutronBooklet}, and the $g$ value of $g_{xy}=1.9$ from our fits to the inelastic spectrum \cite{bdcInelastic} was used.  All calculated intensities were normalized to the (0 0 2) calculated intensity to account for the unknown constant of proportionality.

$\theta$ scans of the (0 0 $L$) Bragg peaks were background subtracted and integrated as a measure of the total peak intensity.  For odd values of $L$, the scans taken at $\mu_0H= 0.5$ T were used as a measure of the background.  For the structural peaks at even values of $L$, a linear fit to the four points furthest from the peak was used to estimate background.  To get a measure of the field-induced magnetic intensity at (0 0 4), the zero-field scan was subtracted from the 0.5-T scan to remove both background and the structural peak signal.  Integrated values were normalized to the measured intensity of the (0 0 2) peak and then divided by the calculated intensity.  

\begin{figure}
\includegraphics[width=8.5cm]{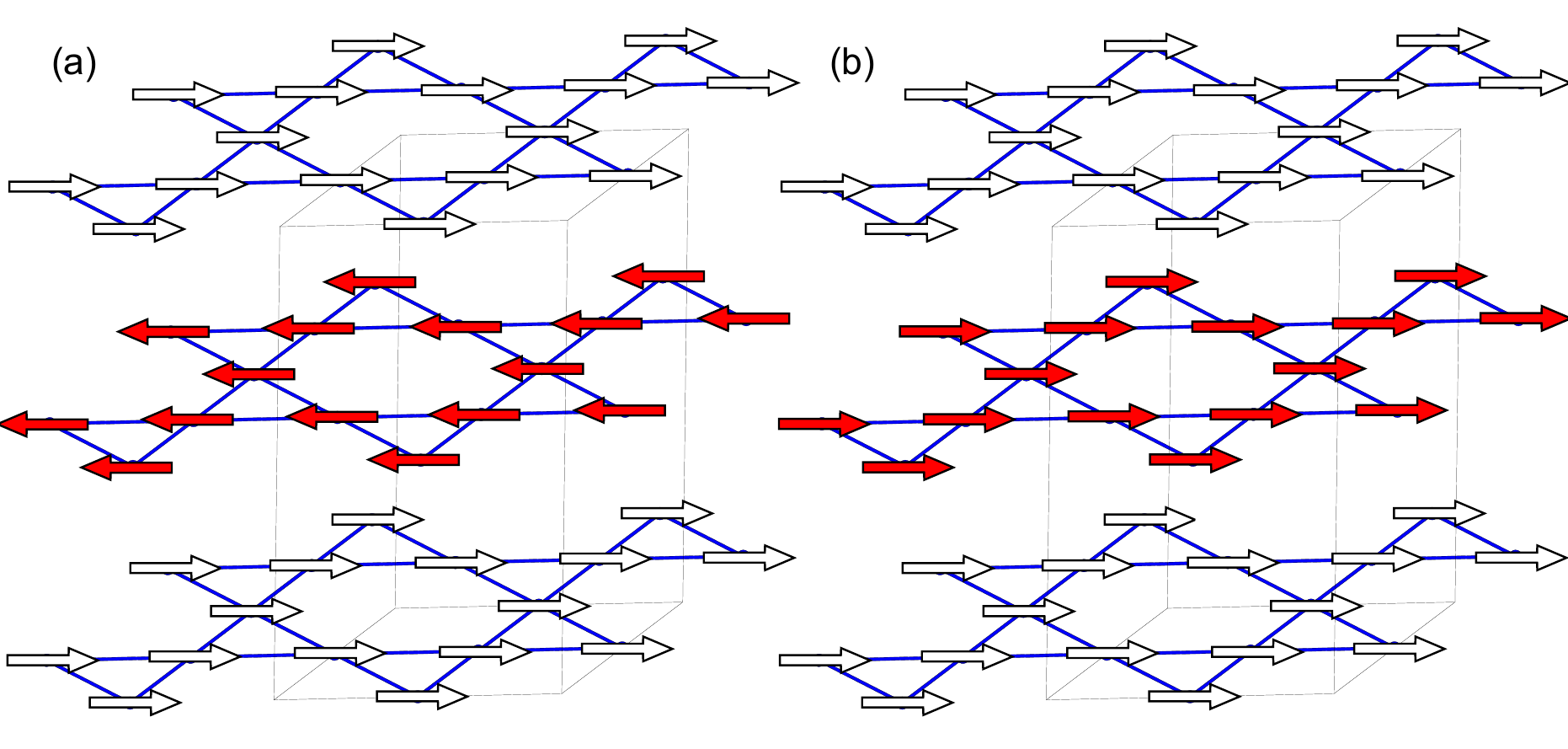} \vspace{0mm}
\caption{Schematic of ground state spin configuration at (a) zero applied field and (b) $\mu_0H\gtrsim 0.05$ T with the field applied parallel to the kagome plane.  The Cu$^{2+}$ ions are ferromagnetically ordered within each kagome plane and constrained to point parallel to the kagome plane.  Neighboring planes are antiferromagnetically ordered at zero field but are easily aligned by a small magnetic field.}
\end{figure}

Figure 6 shows the ratio of measured peak intensity to calculated peak intensity.  The value of 1.0 indicates that the measured and calculated intensities agree.  The structural peaks were included as a check of our vertical divergence calculation, and the field-induced peak at (0 0 4) was included to check that the use of the free Cu$^{2+}$ ion form factor was a reasonable approximation of the true form factor.  The antiferromagnetic peak intensities (red circles) provide a measured value for the ordered moment.  Since the calculated value of the intensity assumed spins were aligned parallel to the kagome plane, the intensities shown in Fig. 6 can be interpreted as the ratio of the square of the measured moment to the square of the full moment:
\begin{equation}
\frac{\mathcal{I}_{measured}}{\mathcal{I}_{calculated}}=\frac{(g_{xy}S_{measured})^2}{(g_{xy}S)^2}
\end{equation} 
From our measurements we can determine a value of the ordered moment of $g_{xy}S_{measured}$ = (0.95 $\pm$ 0.2) $\mu_B$, which suggests that the spins point entirely within the kagome plane.  The stated uncertainty represents one standard deviation.

Figure 7 shows a schematic of the ground state spin configuration, which summarizes the results of our diffraction measurements.  At zero field, spins within each kagome plane are ferromagnetically ordered and point parallel to the kagome plane, while neighboring kagome planes are antiferromagnetically ordered.  Due to the existence of two copper layers per unit cell, the magnetic unit cell is equivalent to the structural unit cell.  A small magnetic field ($\mu_0H \gtrsim 0.05$ T at 70 mK) reorients the spins so that neighboring planes are ferromagnetically ordered.

\subsection{Interplane Coupling}

\begin{figure}
\includegraphics[width=8.5cm]{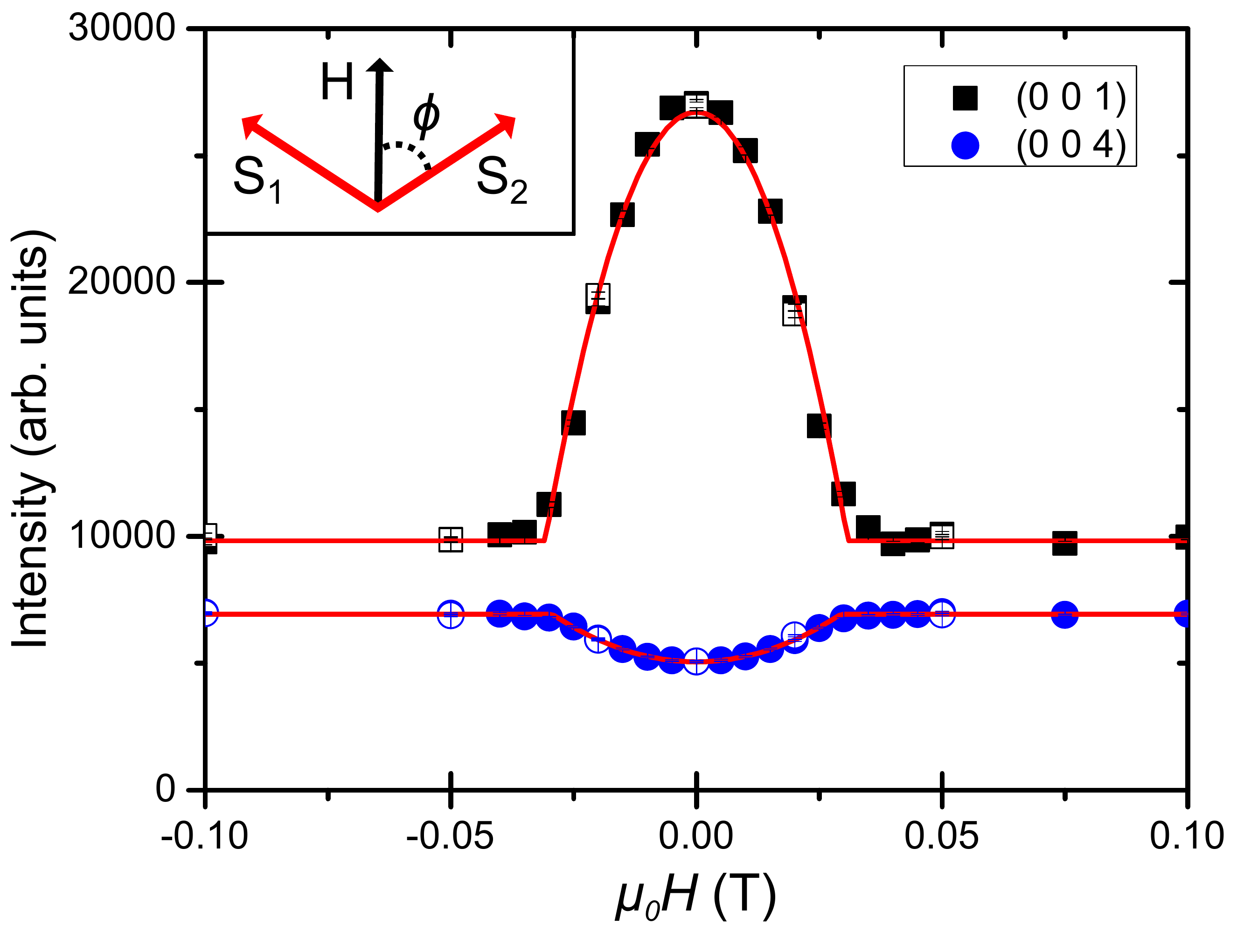} \vspace{0mm}
\caption{Bragg peak intensities plotted as a function of applied magnetic field at $T=70$ mK.  Data measured with field increasing (open) and decreasing (solid) are shown.  Lines are fits as described in the text.  (Inset) Schematic of the two-sublattice model used to fit the data.}
\end{figure}

To determine the interplane magnetic coupling, we examine the field dependence of the magnetic Bragg peak intensities, shown in Fig. 8.  The (0 0 1) peak is fully suppressed by $\mu_0H\approx$ 0.03 T.  The falloff in intensity is symmetric about $H = 0$ and shows no sign of hysteresis.  The peak at (0 0 4) reaches its peak intensity at the same field and is not enhanced further up to $\mu_{0}H$ = 10 T.  

As evidenced by the small fields required to fully polarize the magnetic moments in Cu(1,3-bdc), the antiferromagnetic interplane coupling $J_{AF}$ is much smaller than the ferromagnetic in-plane nearest neighbor coupling $J$.  Therefore, we assume the application of a magnetic field only changes the relative orientation of neighboring planes and does not disturb the ferromagnetic ordering of each plane.  In other words, we treat this system as a 1D chain of antiferromagnetically coupled classical spins in a magnetic field.  The energy of the system is

\begin{equation}
J_{AF}\sum_{\left<i,j\right>}\vec{S}_i\cdot\vec{S}_j+g\mu_B\vec{H}\cdot\sum_i \vec{S}_i
\end{equation}
where $\vec{S}_i$ is the Cu$^{2+}$ spin moment at site $i$ and $\left<i,j\right>$ indicates summation over interplane nearest neighbors.  The system has per-spin energy
\begin{equation}
J_{AF}S^2\cos(2\phi) - g\mu_BHS\cos(\phi)
\end{equation}
where $2\phi$ is the angle between the two sublattices, $S=1/2$, and $g=1.9$.  Thus the ground state configuration has
\begin{equation}
\phi=
\begin{cases}
\cos^{-1}(\frac{g\mu_BH}{4J_{AF}S}) & g\mu_BH<4J_{AF}S\\
                  0                                  &  g\mu_BH\geq4J_{AF}S
\end{cases}
\end{equation}
This configuration is shown schematically in the inset of Fig. 8.  The antiferromagnetic peak at (0 0 1) is due to the antiparallel component of the spins, and therefore its intensity is proportional to $\sin^2\phi$, while the (0 0 4) peak intensity is proportional to $\cos^2\phi$.  Lines in Fig. 8 show fits to the data using these functions, which return a value of $J_{AF}=1.65(4) \mu eV$.  Thus $\left|J_{AF}/J\right|\approx0.003$ and the treatment of the magnetic behavior of Cu(1,3-bdc) as 2D is justified.

\subsection{Magnetic Ordering Transition}

In this section we address the nature of the magnetic ordering transition by comparing the temperature dependence observed in elastic and inelastic neutron scattering measurements.

To examine the critical behavior of the magnetic transition, a temperature scan of the 0-field (0 0 1) peak intensity was performed.  Temperature control of the magnet cryostat became unstable above 1.3 K.  In order to investigate the behavior near the transition temperature of $\approx$1.8 K, we combined three different measurements of the (0 0 1) peak intensity.  First, below 1.3 K measurements of the peak intensity were taken with the temperature stable at a set point.  Second, measurements were performed continuously in 30-s increments while cooling the sample from above 2 K to the base temperature.  For these points a temperature error bar is included which is the difference in temperature between the start and end of the 30-s measurement interval.  We note that this measurement is taken with zero applied field while cooling through the superconducting transition of the aluminum sample holder.  However, the peak intensities obtained from this measurement agree with those from the first measurement, which was performed after application of a field.  Third, a temperature scan of the (0 0 1) peak intensity was performed with the sample in the He-4 cryostat, allowing for measurement around 1.8 K at stable temperatures.  

To compare the intensities of the diffraction measurements taken in the 10-T magnet and the elastic scattering measurements taken in the He-4 cryostat, background signal was estimated by fitting a constant to data points measured at temperatures above 3 K.  After subtracting the background from each data set, the overlapping data point at $T=1.63$ K was scaled to be the same in both data sets.  By combining these three measurements as shown in Fig. 9(a), we were able to examine the behavior of the (0 0 1) peak through the transition temperature.  This combined temperature scan was fit to a power law $I\propto$ (1-$\frac{T}{T_{c}}$)$^{a}$, resulting in $a$ = 0.492(6) and $T_{c}$ = 1.77(2) K.  This value of $T_{c}$ is consistent with the transitions observed in our specific heat measurements and in $\mu$SR \cite{bdcMuon} measurements.

We also examined the temperature dependence of the magnetic field-induced peak at (0 0 4).  A field of magnitude $\mu_{0}H=0.05$ T was applied and the (0 0 4) peak intensity was measured while cooling from 2.4 K to 70 mK.  The field strength of 0.05 T was chosen because it was strong enough to produce the maximum intensity of the (0 0 4) peak at $T=70$ mK.  To isolate the magnetic component of the scattering, the (0 0 4) peak intensity measured at $T=70$ mK and zero applied field was subtracted to remove the contributions from the structural peak and from background.  The field broadens the transition and shifts it to higher temperatures.

\begin{figure}
\includegraphics[width=8.5cm]{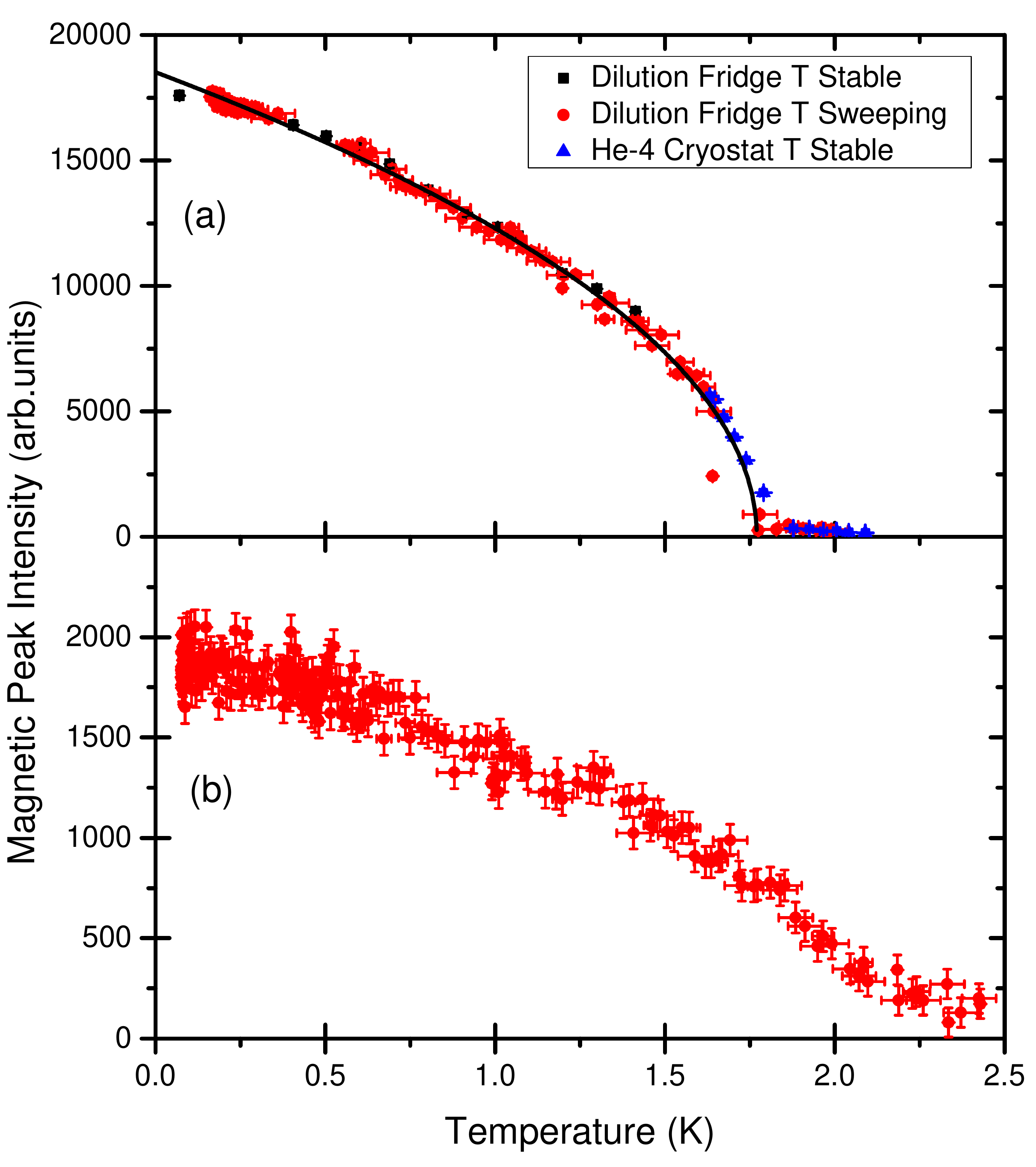} \vspace{0mm}
\caption{\label{Figure1} Magnetic Bragg peak intensities plotted as a function of temperature. (a) (0 0 1) peak intensity with zero applied field. Background was estimated by fitting a constant to high temperature points.  The line is a power law fit as described in the text. (b) (0 0 4) peak intensity with $\mu_{0}H = 0.05$ T.  The (0 0 4) peak intensity with $T=70$ mK and $H=0$ was subtracted to remove the structural peak and background intensities.}
\end{figure}

In the magnetically ordered state, the magnetic excitation spectrum includes a flat mode at energy transfer 1.8 meV \cite{bdcInelastic}.  We investigated the temperature dependence of this flat mode using inelastic scattering measurements of a powder sample of Cu(1,3-bdc).  Figure 10(a) shows the inelastic neutron scattering data integrated over a range of momentum transfers 1.0 $\mathrm{\AA}^{-1}\leq\vec{\left|Q\right|}\leq$ 3.4 $\mathrm{\AA}^{-1}$ at 100 mK and 40 K.  Intensities at different temperatures were normalized by integrating over the measured energy range $-3.7$ meV $\leq\hbar\omega\leq$ 6.1 meV.  We assume the total scattering intensity to be constant over this interval.  The peak at 1.8 meV appears at low temperatures and is due to the non-dispersive magnetic excitation.  The smaller peak seen in the 40 K data around 1.9 meV is a temperature-independent background signal most likely due to scattering from the cryostat.  

\begin{figure}
\includegraphics[width=8.5cm]{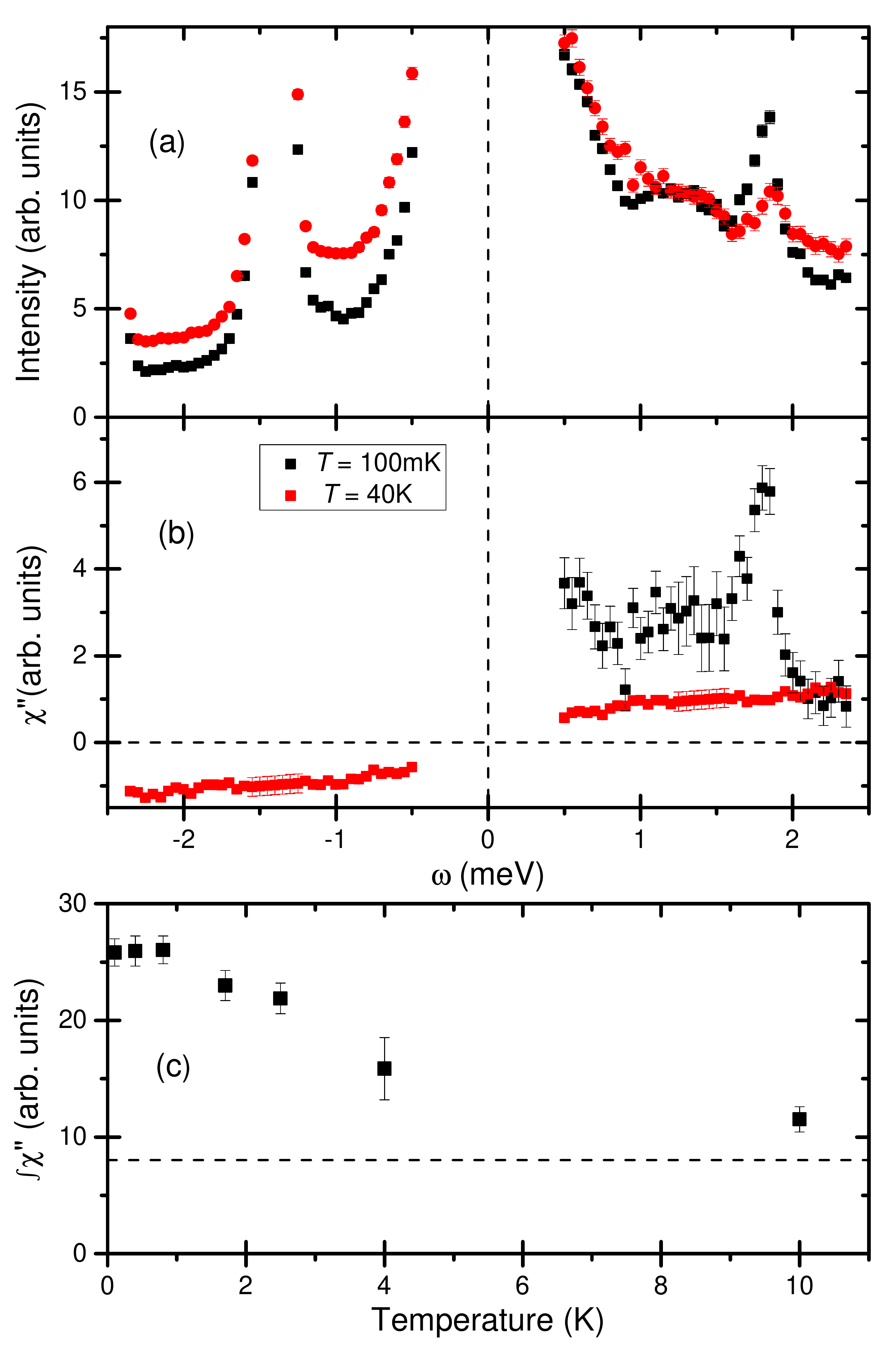} \vspace{0mm}
\caption{(a) Inelastic neutron scattering data measured on a powder sample integrated over momentum transfers 1.0 $\mathrm{\AA}^{-1}\leq\vec{\left|Q\right|}\leq 3.4\;\mathrm{\AA}^{-1}$. (b) $\chi^{\prime\prime}(\omega)$ extracted from the data as described in the text. (c) $\chi^{\prime\prime}(\omega)$, integrated over the energy range 1.5 meV $\leq\omega\leq$ 2.1 meV spanning the low temperature peak, as a function of temperature.  The dashed line indicates the value at $T$ = 40 K.  Vertical error bars throughout this paper represent one standard deviation.}
\end{figure}

The inelastic scattering signal is proportional to the dynamic structure factor $\mathbf{S}(\vec{Q},\omega)$ = [$\mathrm{n}(\omega)$+1]$\chi''$($\vec{Q},\omega$), where $\mathrm{n}(\omega)$ is the Bose occupation factor and $\chi''$($\vec{Q},\omega$) is the imaginary part of the dynamic susceptibility.  To isolate the inelastic signal due to scattering from the sample, we applied the following procedure, following Helton \emph{et al.} \cite{HerbertsmithitePowder}.  For negative energy transfers at low temperatures, the scattered intensity is only background because the scattering from the sample is suppressed by the Bose factor.  Therefore, $\chi''(\omega,T = 40\; \mathrm{K})$ can be calculated by subtracting the intensity measured at 100 mK from that measured at 40 K and dividing by the Bose factor.  Then $\chi''(\omega,T = 40\; \mathrm{K})$ is known for positive energy transfers because $\chi''$($\omega$) is an odd function of $\omega$.  The positive energy transfer background can be calculated by subtracting the calculated signal at 40 K from the measured intensity at 40 K.  Assuming this background is temperature-independent in the range 100 mK to 40 K, this background can be subtracted from the intensities measured at other temperatures to arrive at $\mathbf{S}$($\omega$,$T$).  $\chi''$($\omega$,$T$) is then calculated by dividing by the Bose factor.

Figure 10(b) shows $\chi''$($\omega,T$ = 40 K).  Points at small energy transfer, $\left|\omega\right|<0.5\;\mathrm{meV}$ were removed because at these energy transfers the intensity at 100 mK is not only background, but also includes a contribution from the elastic line due to the instrumental resolution.  The peak in the scattered intensity at $\omega\approx$ $-1.4$ meV [see Fig. 10(a)] is a known background signal most likely due to scattering from aluminum windows, and its effects are not fully canceled out by subtracting the two data sets.  Points in the range 1.25 meV $\leq\left|\omega\right|\leq$ 1.55 meV were removed and replaced by fitting a smooth function to the remaining data points.  $\chi''$($\omega,T$ = 40 K) was then used to calculate  $\chi''$($\omega$) for the other measured temperatures.  $\chi''(\omega,T = 100\;\mathrm{mK})$ is also shown in Fig. 10(b).

$\chi''$($\omega$) was integrated over the range 1.5 meV $\leq\left|\omega\right|\leq$ 2.1 meV for each temperature to get a measure of the flat mode intensity.  The results are shown in Fig. 10(c).  Significant spectral weight remains in this energy transfer range well above the 3D transition temperature of 1.77 K seen in the temperature scan of the magnetic Bragg peaks [Fig. 9(a)].  This suggests the existence of 2D correlations within the kagome planes at temperatures above the ordering transition temperature.  This is supported by our observation that a significant fraction of the magnetic entropy is lost in the temperature range 2 K $<T<$ 5 K, as discussed in Sec. III(A).

\section{Conclusions}

We performed detailed thermodynamic and neutron scattering measurements of the 2D TMI material Cu(1,3-bdc).  We examine the magnetic structure in the low-temperature ordered state and show that under zero applied field spins point parallel to the kagome plane.  Magnetization measurements reveal that a small magnetic field can fully polarize the spins in any direction, although they are most easily polarized parallel to the kagome plane.  We use field-dependent neutron scattering measurements to deduce the antiferromagnetic interplane coupling, and show that it is $\approx$0.3 \% of the ferromagnetic in-plane nearest-neighbor coupling.  This confirms that the treatment of the magnetic behavior of Cu(1,3-bdc) as 2D is justified.  Specific heat and neutron scattering measurements show a clear 3D magnetic ordering transition at $T_c=1.77$ K but also reveal significant magnetic correlations at much higher temperatures, consistent with 2D behavior.  At least 40 \% of the spin entropy is lost at temperatures above $T_c$, where significant spectral weight also remains in the topologically nontrivial flat magnon band.  Our results confirm that Cu(1,3-bdc) is an ideal test material for examining 2D spin physics with a simple Hamiltonian and provide a more complete understanding of the magnetic ordering in Cu(1,3-bdc) that gives rise to the TMI state.

\begin{acknowledgments}
This work was supported by the U.S. Department of Energy (DOE), Office of Science, Basic Energy Sciences, Materials Sciences and Engineering Division, under Contract No. DE-AC02-76SF00515.  D. E. F. acknowledges support from the National Science Foundation, Grant No. CHE 1041863.  We acknowledge the support of the National Institute of Standards and Technology, U.S. Department of Commerce, in providing the SPINS neutron facility used in this work.  Experiments on Iris at the ISIS Pulsed Neutron and Muon Source were supported by a beamtime allocation from the Science and Technology Facilities Council.  The identification of any commercial product or trade name does not imply endorsement or recommendation by the National Institute of Standards and Technology.
\end{acknowledgments}

\bibliography{cubdcelastic}{}

\end{document}